\documentclass[aps,pra,preprint,superscriptaddress,twocolumn, 10pt]{revtex4}
\usepackage{graphicx}
\usepackage{epsfig}
\usepackage{epstopdf}
\usepackage{subfigure}
\usepackage{appendix}
\usepackage{relsize}
\usepackage{amsmath,amssymb}
\begin{document}
\title{Probing vacuum-induced coherence via magneto-optical rotation in molecular systems}
\author{Pardeep Kumar}
\email[]{pradeep.kumar@iitrpr.ac.in}
\affiliation{Department of Physics, Indian Institute of Technology Ropar, Rupnagar, Punjab 140001, India}
\author{Bimalendu Deb}
\affiliation{Department of Material Science and Raman Center for Atomic, Molecular and Optical Sciences, Indian Association for the Cultivation of Science, Jadavpur, Kolkata 700032, India}
\author{Shubhrangshu Dasgupta}
\affiliation{Department of Physics, Indian Institute of Technology Ropar, Rupnagar, Punjab 140001, India}

\date{\today}

\begin{abstract}
We investigate theoretically the effects of vacuum-induced coherence (VIC)  on magneto-optical rotation (MOR). We carry out a model study to show that VIC in the presence of a control laser and a magnetic field can lead to large enhancement in the rotation of the plane of polarization of a linearly polarized weak laser with vanishing circular dichroism. This effect can be realized in cold molecular gases and may be used as a sensitive probe for VIC.  Such a large MOR angle can also  be used to detect weak magnetic field with large measurement sensitivity.
\end{abstract}


\maketitle

\section{Introduction}
Many prodigious phenomena in atomic and molecular physics are the consequences of quantum coherence and interference \cite{ficek2007}. A pronounced coherence phenomenon, called \textit{vacuum-induced coherence} (VIC), arises due to the quantum interference between the spontaneous emission pathways from the excited doublet  to a common ground state \cite{agarwal1974}. The coupling between decay pathways via the same continuum of vacuum states of electromagnetic fields create these interfering channels. Recently, numerous theoretical studies have reported fascinating applications of such quantum interference \cite{scully1989, syzhu1995, harris1989, knight1982, zhou1996, zhou1997, knight1998, keitel1992, zhu1995, zhu1996, swain1997, lee1997, keitel1999, menon1998, menon1999}. More than two decades ago, it was shown that VIC can lead to the cancellation of spontaneous emission from the two nondegenerate upper levels of a three level system \cite{scully1989}. Quantum interference effects on the spectrum of emission from two upper levels driven by a coherent field was studied by Zhu \cite{syzhu1995}. In a model study, the elimination of spectral line in the spontaneous emission spectrum was demonstrated by Zhu and Scully \cite{zhu1996}. In an analogous problem, the suppression of autoionization in a three-level system with two degenerate upper levels was shown by Harris \cite{harris1989}. In all such theoretical studies on quantum interference, the common underlying basis is the interaction of discrete levels with a continuum of states.  One of the stringent requirement for the VIC to occur is the \textit{nonorthogonality} of the  dipole matrix elements of the two transitions from degenerate or quasi-degenerate excited states \cite{imamoglu1989} to the common ground state. It  is very difficult, if not impossible, to fulfill this condition in atomic systems. However, it is shown in \cite{agarwal2000,li2001} that this condition can be bypassed by placing the atom in \textit{anisotropic vacuum}. The simulation of quantum interference in atomic systems without near-degenerate levels  has also been suggested \cite{ficek2007}. Moreover, possible realizations of VIC have been proposed in ions \cite{kiffner2006,das2008}, semiconductor quantum-well \cite{faist1997}, quantum dots \cite{dutt2005,sophia2005} and in  M\"{o}ssbauer nuclei \cite{heeg2013}. Despite these attempts, a clear manifestation of VIC in atomic systems  still remains elusive.

On the other hand, the situation is entirely  different in cold molecules which appear to be quite promising systems for exploration of VIC. In fact, the VIC can appear naturally  in molecular systems \cite{das2012}, as it is possible to identify non-orthogonal dipole transitions in molecules unlike in atoms. Such non-orthogonal transitions in a diatomic molecule arises due to the coupling of the rotation of molecular axis with molecular electronic angular momentum. It is possible to find diatomic molecules having suitable V-type three level structures, with both upper levels being either degenerate or non-degenerate for exploring VIC. Consider a molecule initially prepared in a particular vibrational level $v_{0}$ and rotational level $J=0$ in spin-singlet ($S=0$) electronic ground state. Let this initial state be represented by $|0\rangle=|v_{0},J=0,m_{J}=0;S=0,L=0,M_{L}=0\rangle$, where $L$ and $M_{L}$ denote the molecular electronic orbital angular momentum and its projection onto the the internuclear axis  which is molecule-fixed $z$-axis. Let us consider laser transitions from this state to two excited molecular states $|1\rangle=|v_{1},J=1,m_{J}=1;S=0,L=1,M_{L}\rangle$ and  $|2\rangle=|v_{2},J=1,m_{J}=1;S=0,L=1,M_{L}\rangle$ which have the same rotational and the same molecular electronic states. $v_{1}$ and $v_{2}$ represent the vibrational quantum numbers of the two excited states. Now, in the first approximation, when higher order spin-orbit, spin-rotation and orbit-rotation interactions are neglected, one of the good molecular quantum numbers in Hund's case (a) and (b) is $\Lambda=|M_{L}|$. Now, if we choose $M_{L}=0$, that is, both the excited levels belonging to $^{1}\Sigma$ states, $v_{1}$ and $v_{2}$ should be necessarily different and so the two excited states are non-degenerate. On the other hand, if we select $M_{L}=\pm 1$ or equivalently, $\Lambda=1$, that is both the excited levels being in the $^{1}\Pi$ states, $v_{1}$ and $v_{2}$ may or may not be equal. If it is chosen that $v_{1}=v_{2}$, then the two states with $M_{L}=\pm 1=\pm \Lambda$ are degenerate. The degeneracy will be lifted by $\Lambda$-doubling effect \cite{landau1977}. In this particular case, $\Lambda$-doubling will occur due to orbit-rotation coupling. In both the non-degenerate ($v_{1}\neq v_{2}$) and the degenerate ($v_{1}=v_{2}$) cases, the molecular transition dipole moments are parallel. The non-degenerate case with $M_{L}=0$ and $v_{1}\neq v_{2}$ satisfy exactly the same condition as considered by Imamoglu \cite{imamoglu1989} in a model three-level atomic system  for realization of VIC effects about twenty six years ago. Probably, it is difficult to find such an atomic system having two upper levels with same electronic angular momentum states. But, molecular systems, particularly the emerging area of cold molecules presents a promising perspective for realization of VIC and related effects. In the particular scheme of molecular transitions we consider here,  the two transitions occur between the same rotational and electronic levels. Therefore, the electronic part of the dipole moment matrix elements will be the same, as the two dipole moments are parallel. In the non-degenerate case, these matrix elements differ only slightly by the Franck-Condon (FC) overlap integrals for the two vibrational states which may be taken as the two nearby even or odd vibrational levels with large vibrational quantum numbers  so that the energy spacing between them is small. 

Usually, population oscillation in the excited vibrational states is understood as an effect of VIC and thus is used to detect VIC in experiments. As the population in the excited state occurs due to \textit{absorption} of the probe field, it is the \textit{imaginary part} of the susceptibility of the medium, that plays the most important role in  such detection. In this work, we present a different strategy - we show how the \textit{dispersion} of the probe field (or equivalently, the \textit{real part} of the susceptibility) can be manipulated to obtain a measurable signature of VIC. Precisely speaking, we explore a way to probe VIC in  molecules by observing its influence on magneto-optical rotation (MOR). MOR refers to the rotation of the plane of polarization of light, while propagating through the medium in the presence of magnetic field.
It occurs essentially due to birefringence or dichroism induced in the medium by the applied magnetic field.

 The angle of MOR of a linearly polarized weak field propagating through a medium with negligible absorption is given by
\begin{equation}
\Theta=\pi k_{p}L\mbox{Re}\left(\chi_{-}-\chi_{+}\right)\;.
\label{eq1}
\end{equation}
where, $\chi_{\pm}$ represent the susceptibilities of the medium corresponding to right (+) and left (-) circular polarization of the probe field, respectively, $k_{p}$ is the propagation constant of the weak field and $L$ is the length of the medium. MOR angle can be enhanced by creating large anisotropy in the medium using a high magnetic field. Furthermore, the enhancement can also be accomplished by  a coherent control field \cite{patnaik2000,patnaik2001,vasant2008}. A nice review on MOR and its applications can be found  in \cite{budker2002}.
 
 Here we show that in the presence of VIC, the angle of rotation of the plane of polarization of linearly polarized light propagating through a medium of cold molecules can be significantly large at resonance. Further, the angle of rotation as large as $180^{\circ}$ can be achieved by employing a control field and a magnetic field. For this purpose a medium of cold molecules \cite{krem2009} is useful, as it is possible to suppress thermal fluctuations so that low lying rotational states can be accessed for the existence of VIC . The  large MOR angle so obtained  can be employed as a tool for the sensitive detection of very weak magnetic field \cite{budker1998}.  A small change in magnetic field can lead to large change in the MOR angle, and thereby to large measurement sensitivity \cite{budker2007,kominis2003,fleischhauer1994,lee1998,petrosyan2004}. 

The organization of the paper is as follows. In Sec. II we describe the theoretical model with relevant density matrix equations including VIC. In Sec. III we delineate a realistic molecular system where VIC can be perceived. We discuss in Sec. IV that the VIC in the presence of the control and magnetic  field leads to large MOR angle. The main results of the paper are presented in this section. In Sec. V we propose a magnetometer based on VIC for the sensitive detection of the magnetic field. Sec. VI, describe the conclusions and discussion on the possible experimental realization for the observation of VIC in cold molecular systems.

\section{Relation between polarization rotation and VIC}
\subsection{The model}
\begin{figure}[!ht]
\begin{center}
\includegraphics[scale=0.28]{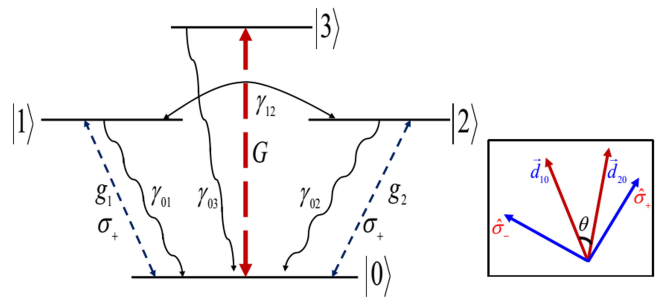}
\end{center}
\caption{(Color online)  Simplified schematic of the energy-level diagram. The levels $|1\rangle$ and $|2\rangle$ are coupled to ground state $|0\rangle$ by $\sigma_{+}$ component of the probe field. The VIC arises between $|1\rangle$ and $|2\rangle$ from the quantum interference of the two spontaneous decay channels $|1\rangle\rightarrow|0\rangle$ and $|2\rangle\rightarrow|0\rangle$.  The arrangement of the field polarizations and the dipole moments is shown in the box.}
\label{fig1}
\end{figure}
To understand the basic idea of enhancement of polarization rotation in presence of VIC, we first consider a generic energy level scheme [see Fig. \ref{fig1}]. For simplicity, we consider that the system comprises of two near-degenerate excited molecular states $\left( |1\rangle \right)$ and $\left( |2\rangle \right)$   coupled to a ground rovibrational state $\left( |0\rangle \right)$. Cold molecules can be prepared in such states by two-photon Raman photoassociation of cold atoms \cite{thorsheim1987,jones2006}. A linearly polarized weak probe field  $\vec{E}_{p}=\hat{x}\varepsilon_{p}e^{-i\left(\omega_{p}t-k_{p}z\right)}+c.c.$ of frequency $\omega_{p}$ and propagation constant $k_{p}$ is applied to drive the transitions  as shown in Fig. \ref{fig1}. The $\sigma_{+}$ component of the probe field couples to the transitions $|0\rangle\leftrightarrow|1\rangle$ and $|0\rangle\leftrightarrow|2\rangle$, with the respective Rabi frequencies $2g_1$ and $2g_2$, following the dipole selection rules \cite{spano2001}, while the $\sigma_-$ component does not couple to these transitions.
A $\pi$-polarized control field $\vec{E}_{c}=\hat{z}\varepsilon_{c}e^{-i\omega_{c}t}+c.c.$ of Rabi frequency $2G$ couples the ground state $|0\rangle$ to an auxiliary level $|3\rangle$, where $\omega_{c}$ is the frequency of the control field. The coherence created by this field between levels $|0\rangle$ and $|3\rangle$ causes high refractive index in the medium without absorption \cite{scully1991,fleischhauer1992,rathe1993}.

The Hamiltonian of this system in the dipole approximation can be written as
\begin{equation}
\begin{array}{c}
\hat{H}=\hbar\sum\limits_{i=1}^{3}\omega_{i0}|i\rangle\langle i|-\sum\limits_{i=1}^{2}\left[\left(\vec{d}_{i0}|i\rangle\langle 0|+h.c. \right)\cdot\vec{E}_{p}\right]\\
-\left[\left(\vec{d}_{30}|3\rangle\langle 0|+h.c.\right)\cdot\vec{E}_{c}\right]\;.\\
\end{array}
\label{eq2}
\end{equation}
Here zero of the energy is defined at the level $|0\rangle$, $\hbar \omega_{ij}$ is
the energy difference between the levels $|i\rangle$ and $|j\rangle$ and $\vec{d}_{ij}$ is the transition dipole
moment matrix element of the transition $|i\rangle\leftrightarrow |j\rangle$.
The dynamical evolution of the system can be described by Markovian master equation:
\begin{equation}
\begin{array}{lll}
\dot{\rho}&=&-\frac{i}{\hbar}\left[H,\rho\right]+\frac{1}{2}\sum\limits_{i=1}^{3}\gamma_{0i}\left(2A_{0i}\rho
A_{i0}-A_{ii}\rho-\rho A_{ii}\right)\\
& &+ \frac{1}{2} \gamma_{12}\sum\limits_{\begin{array}{c}i,j=1\\i\neq j\end{array}}^{2}\left(2A_{0i}\rho
A_{j0}-A_{ji}\rho-\rho A_{ji}\right)\;.\\
\end{array}
\label{eq3}
\end{equation}
where, $\gamma_{0i}$ is the spontaneous emission rate from the excited state $|i\rangle(i=1,2,3)$
to the ground level $|0\rangle$. Here, the term $\gamma_{12}=\sqrt{\gamma_{01}\gamma_{02}}\cos\theta$ arises due to  VIC, that  results from the cross-talk between two decay paths $|1\rangle\leftrightarrow|0\rangle$ and
$|2\rangle\leftrightarrow|0\rangle$. Here $\cos\theta=\frac{\vec{d}_{10}\cdot\vec{d}_{20}}{|\vec{d}_{10}||\vec{d}_{20}|}$ with $\theta$ being the angle between dipole moment elements $\vec{d}_{10}$ and $\vec{d}_{20}$ [see Fig. \ref{fig1}]. Clearly, for orthogonal dipole moments, $\theta=\pi/2$ and the VIC does not exist between the excited states ($\gamma_{12}=0$). The term $A_{ji}=|j\rangle\langle i|$ represents the population operator  for $j=i$ and a dipole transition operator for $j\neq i$.

To obtain the steady state solutions of density matrix, we expand the density matrix elements to first order in the Rabi frequencies of the probe field, $g_{1}$ and $g_2$, and to all orders in the control field Rabi frequency $G$. The detailed solutions are given in the Appendix A.

\subsection{Relation with VIC}

The zeroth order coherence between the relevant energy levels, in absence of the probe field, are related to the populations $\left(\tilde{\rho}_{ii}^{\left(0\right)}\right)$
\begin{eqnarray}
\tilde{\rho}_{12}^{(0)}&=&\frac{i\gamma_{12}\left(\tilde{\rho}_{11}^{(0)}+\tilde{\rho}_{22}^{(0)}\right)}{2\left(\Delta_{10}^{\ast}-\Delta_{20}\right)}\;,
\label{eq4}\\
\tilde{\rho}_{03}^{(0)} &=& -\frac{iG^{\ast}\left(\tilde{\rho}_{00}^{(0)}-\tilde{\rho}_{33}^{(0)}\right)}{\Delta_{30}^{\ast}}\;.
\label{eq5}
\end{eqnarray}
where, $\Delta_{kj} = \delta_{kj} + i\Gamma_{kj}~(k=1,2,3 ~\mbox{\&} ~ j=0)$, $ \delta_{kj}$ and $\Gamma_{kj}$ are the detuning and  the dephasing rate of coherence, respectively, as defined in the Appendix A. It is obvious from Eq. ({\ref{eq4}}) that vacuum induced coherence exists between levels $|1\rangle$ and $|2\rangle$ even in the absence of any control field provided the populations in either of the levels or in both the levels are nonzero, while the coherence between levels $|0\rangle$ and $|3\rangle$ arises due to the applied control field.

In presence of the weak probe field, the coherence $\tilde{\rho}_{j0}^{+} $  between
$|0\rangle\leftrightarrow|j\rangle$ ($j=1,2$) can be written as
\begin{equation}
\tilde{\rho}_{j0}^{+} = g_{1}\tilde{\rho}_{j0}^{\prime(+1)}+g_{2}\tilde{\rho}_{j0}^{\prime(-1)}\;.
\label{eq6}
\end{equation}
As seen in Eq. (\ref{eq3a}) in the Appendix A, $\tilde{\rho}_{j0}^{\prime(+1)}$ depends upon the coherence [Eq. (\ref{eq5})] induced by the control field. Clearly, by opening up a transition using a control field, one can create a situation, in which the $\sigma_+$ polarization component can exhibit large coherence. This is the reminiscence of the idea of enhancement of refractive index, originally proposed in \cite{scully1991,fleischhauer1992,rathe1993}.

The total first order coherence for the $\sigma_{+}$ component of the probe field can then be expressed as
\begin{equation}
\tilde{\rho}_{+} = \tilde{\rho}_{10}^{+} + \tilde{\rho}_{20}^{+}\;.
\label{eq7}
\end{equation}
This indicates that the coherence can be further influenced by the VIC between the excited levels $|1\rangle$ and $|2\rangle$.  The susceptibility of the medium for $\sigma_{+}$ component is proportional to the relevant coherence, as given by \cite{niharika2012,zhou2009}
\begin{equation}
\chi_{+} =\left(\frac{N|\vec{d}_{+}|^{2}}{\hbar\gamma}\right)\tilde{\rho}_{+}\;,\label{eq8}\\
\end{equation}
where, we have assumed that $\vec{d}_{+}=\vec{d}_{10}\approx\vec{d}_{20}$ and $N$ is number density of the medium.

In this way, the susceptibility $\chi_+$ of the medium can be manipulated by using a control field and VIC, for the $\sigma_+$ polarization component of the probe field, while the coherence $\chi_-$ for the $\sigma_-$  component remains unchanged. This is because the $\sigma_-$ component does not interact with the system of Fig. \ref{fig1}. The difference in their coherences leads to the polarization rotation.

The corresponding rotation angle $\Theta\propto \mbox{Re}\left({\chi}_{-}-{\chi}_{+}\right)$ [see Eq. (\ref{eq1})] implies that polarization rotation of the field at the output increases with increase in the difference in the refractive indices (circular birefringence) of the circular components of the probe field. This control can be made by the presence of the control field and the existence of VIC into the system. This suggests that for a given control field, VIC can be detected and quantified by the amount of rotation angle.

\section{A realistic molecular system}
\begin{figure}[ht!]
\includegraphics[scale=0.09]{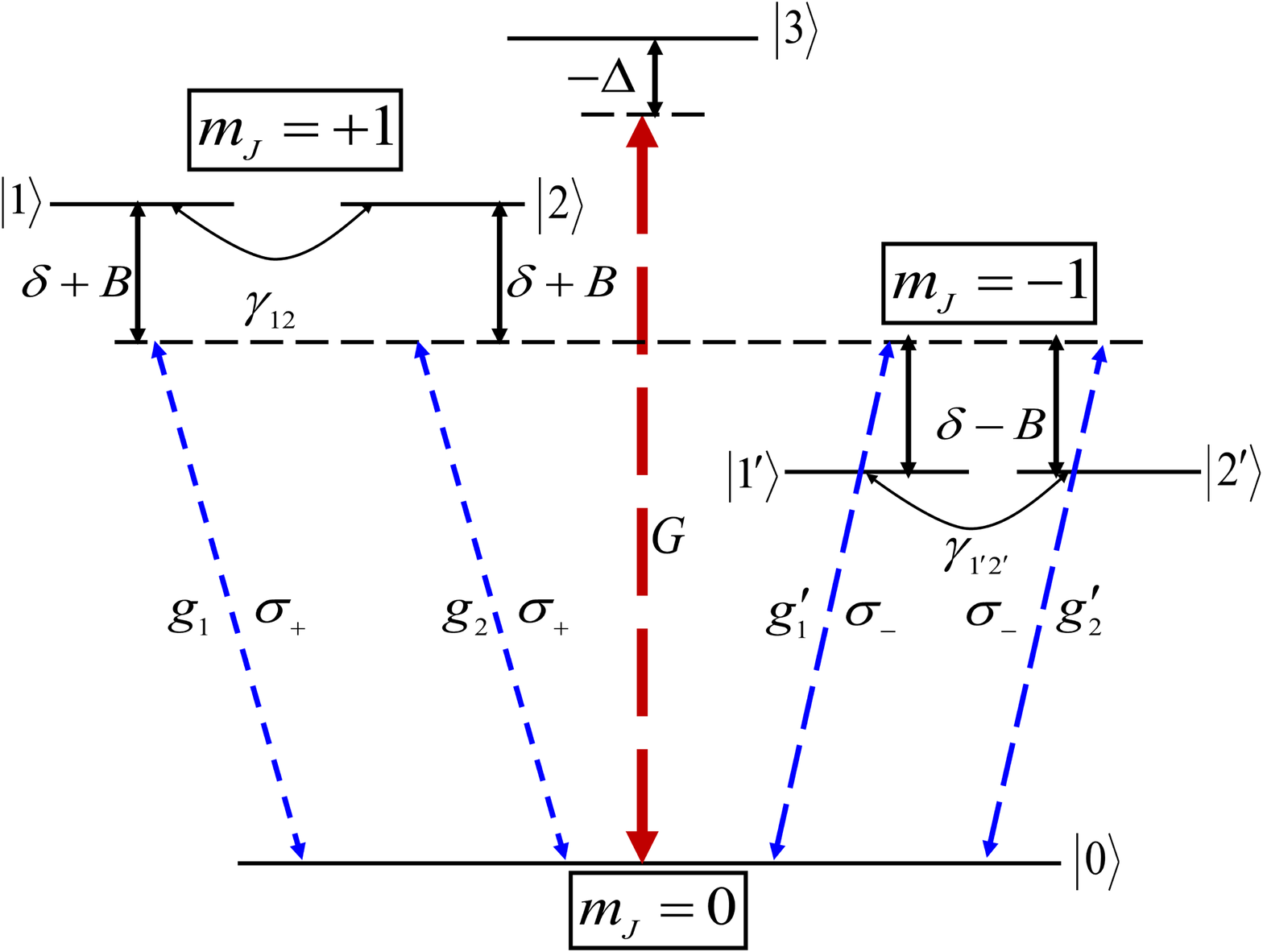}
\caption{ (Color online) Schematic diagram with relevant energy levels. The excited states $|1\rangle$ and $|2\rangle$ $\left(|1^{\prime}\rangle~ \mbox{and}~ |2^{\prime}\rangle\right)$ are the rovibrational levels close to dissociation limit having same $J(=1)$ and $m_{J}$ such that VIC exists between them (see text in Sec. I). The $\sigma_{+}$ ($\sigma_{-}$) component of a $\hat{x}$-polarized probe field drive the transitions $|0\rangle\leftrightarrow|1\rangle$ and $|0\rangle\leftrightarrow|2\rangle$ ($|0\rangle\leftrightarrow|1^{'}\rangle$ and $|0\rangle\leftrightarrow|2^{'}\rangle$), respectively. A $\pi$-polarized control field couples another level $|3\rangle$ to $|0\rangle$. The degeneracy of the excited states is removed by applying an axial magnetic field.  }
\label{fig2}
\end{figure}
Here we discuss how this model can be realized in a molecular system. For VIC to occur, the spacing between the two excited vibrational states should be less than spontaneous linewidth. In the case of near-degenerate $\Lambda$-doubling case, it is easy to fulfill this condition by choosing spin-singlet $\Sigma\leftrightarrow\Pi$ transitions, since the lifting of $\Lambda$-degeneracy due to the orbit-rotation coupling would be very small \cite{landau1977}. In non-degenerate case $(v_{1}\neq v_{2})$, in order to have small spacing between the two excited states, the vibrational levels should be highly excited and lie close to dissociation threshold. Such states will be stable only at a very low temperature. Now, accessing such excited states by optical dipole transitions from a deeply-bound vibrational level in the singlet ground-state molecular potential will be difficult due to extremely poor Franck-Condon factor. However, with recent advances in the production of Feshbach molecules \cite{kohler2006,chin2010} and the cold molecules in the triplet ground-state potentials \cite{lang2008}, excited vibrational levels with appropriate spacing as required for VIC seem to be accessible by transitions from a molecular state in the triplet ground-state potential. For example, RbK \cite{danzl2008,danzl2010} and Cs$_{2}$ \cite{ni2008} in the triplet ground-state in the ultracold temperature regime have been produced. However, for the sake simplicity, we here consider only the near-degenerate case with $v_{1}=v_{2}$.

We now focus on our model configuration as shown in Fig. \ref{fig2}. The $\sigma_-$ component of the probe field interacts with the $(|0\rangle,|1'\rangle, |2'\rangle)$ manifold with the relevant Rabi frequencies $2g_1'$ and $2g_2'$, in which the levels $(|1'\rangle, |2'\rangle)$ can be chosen as near-degenerate rovibrational states with the same quantum numbers $J=1$ and $m_J=-1$. The $\sigma_+$ component couples the excited states $|1\rangle$ and $|2\rangle$ ($m_J=+1$) with the ground state $|0\rangle$. The anisotropy is created by applying a weak magnetic field of strength $B$, such that the excited states $|1\rangle$ and $|2\rangle$ $\left(|1^{\prime}\rangle~ \mbox{and}~ |2^{\prime}\rangle\right)$ are simultaneously Zeeman-shifted by an amount $\delta+B$ ($\delta-B$). Both the manifolds share a common $\pi$-polarized control field with Rabi frequency $2G$, driving the $|0\rangle\leftrightarrow |3\rangle$ transition with a detuning $\Delta$. Clearly, as discussed in the previous section, both the polarization components exhibit enhanced refractive index, however at different frequencies, in presence of the magnetic field.

In this case, the total coherence of the $\sigma_\pm$ components can be written as, in analogy of Eq. (\ref{eq7}),
\begin{equation}
\tilde{\rho}_{\pm} = \tilde{\rho}_{10}^{\pm} + \tilde{\rho}_{20}^{\pm}\;,
\label{eq9}
\end{equation}
where,
\begin{eqnarray}
\tilde{\rho}_{j0}^{+}& =& g_{1}\tilde{\rho}_{j0}^{\prime(+1)}+g_{2}\tilde{\rho}_{j0}^{\prime(-1)}\;,\label{eq10}\\
\tilde{\rho}_{j0}^{-}& =& g'_{1}\tilde{\rho}_{j0}^{\prime(+1)}+g'_{2}\tilde{\rho}_{j0}^{\prime(-1)}\;.\label{eq11}
\label{eq7a}
\end{eqnarray}
The expressions of $\tilde{\rho}_{j0}^{\prime(\pm 1)}$ $(j=1,2)$ are given in the Appendix A.
The susceptibilities of the medium for two polarization components thus become
\begin{equation}
\chi_{\pm} =\left(\frac{N|\vec{d}_{\pm}|^{2}}{\hbar\gamma}\right)\tilde{\rho}_{\pm}\;.\label{eq12}
\end{equation}
The magneto-optical rotation angle can then be written as
\begin{equation}
\Theta=\mathcal{C}~\mbox{Re}\left(\tilde{\rho}_{-}-\tilde{\rho}_{+}\right)\;,
\label{eq13}
\end{equation}
where, $\mathcal{C}=\pi k_{p}L\left(\frac{N|\vec{d}|^{2}}{\hbar \gamma}\right)$ is a constant attributed to the system.
Here, we have assumed $|\vec{d}|=|\vec{d}_{-}|=|\vec{d}_{+}|$.

\section{Results}
\subsection{The effect of control field}
\begin{figure}[!ht]
\begin{center}
\begin{tabular}{lll}
 \includegraphics[scale=0.4]{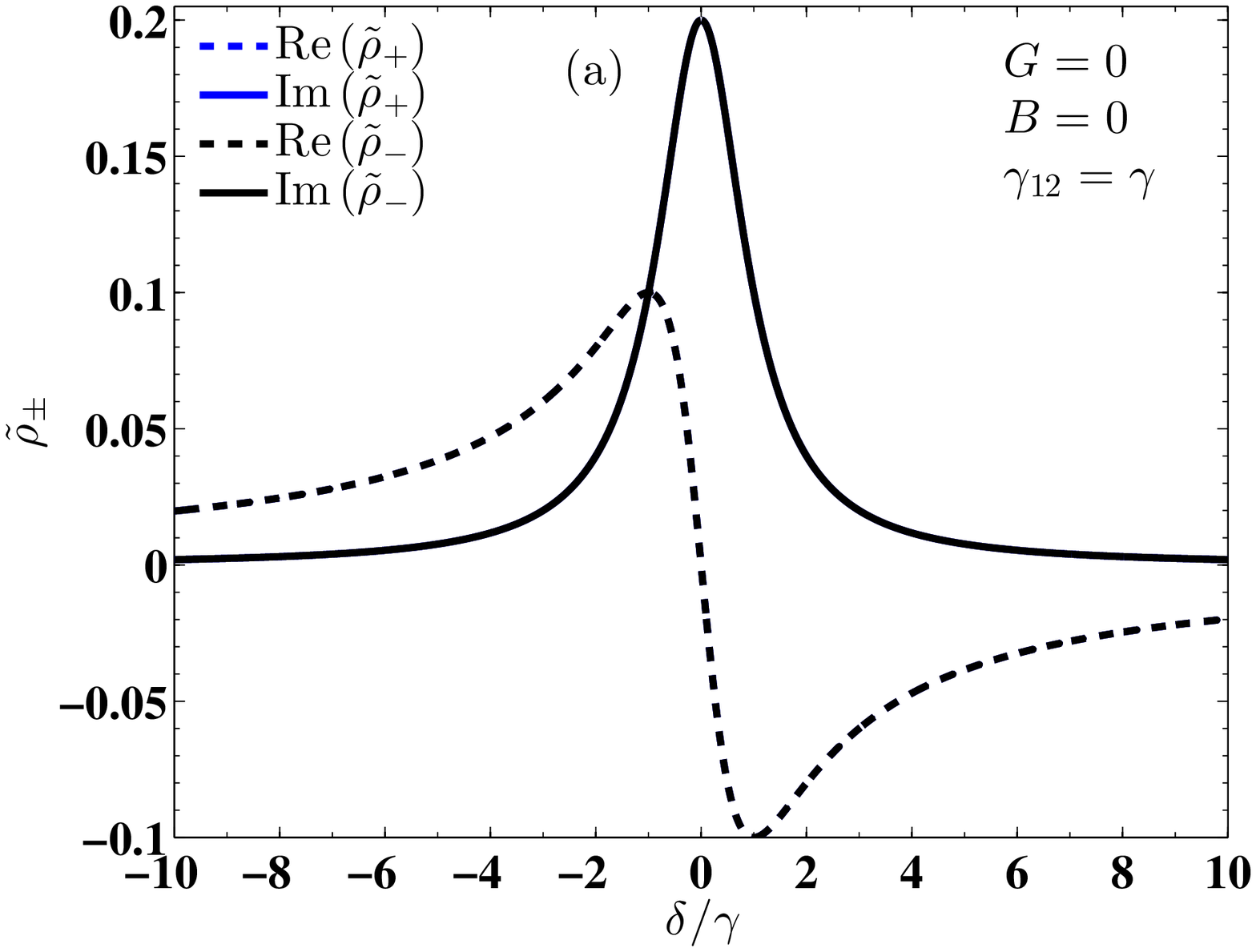}\\
\includegraphics[scale=0.4]{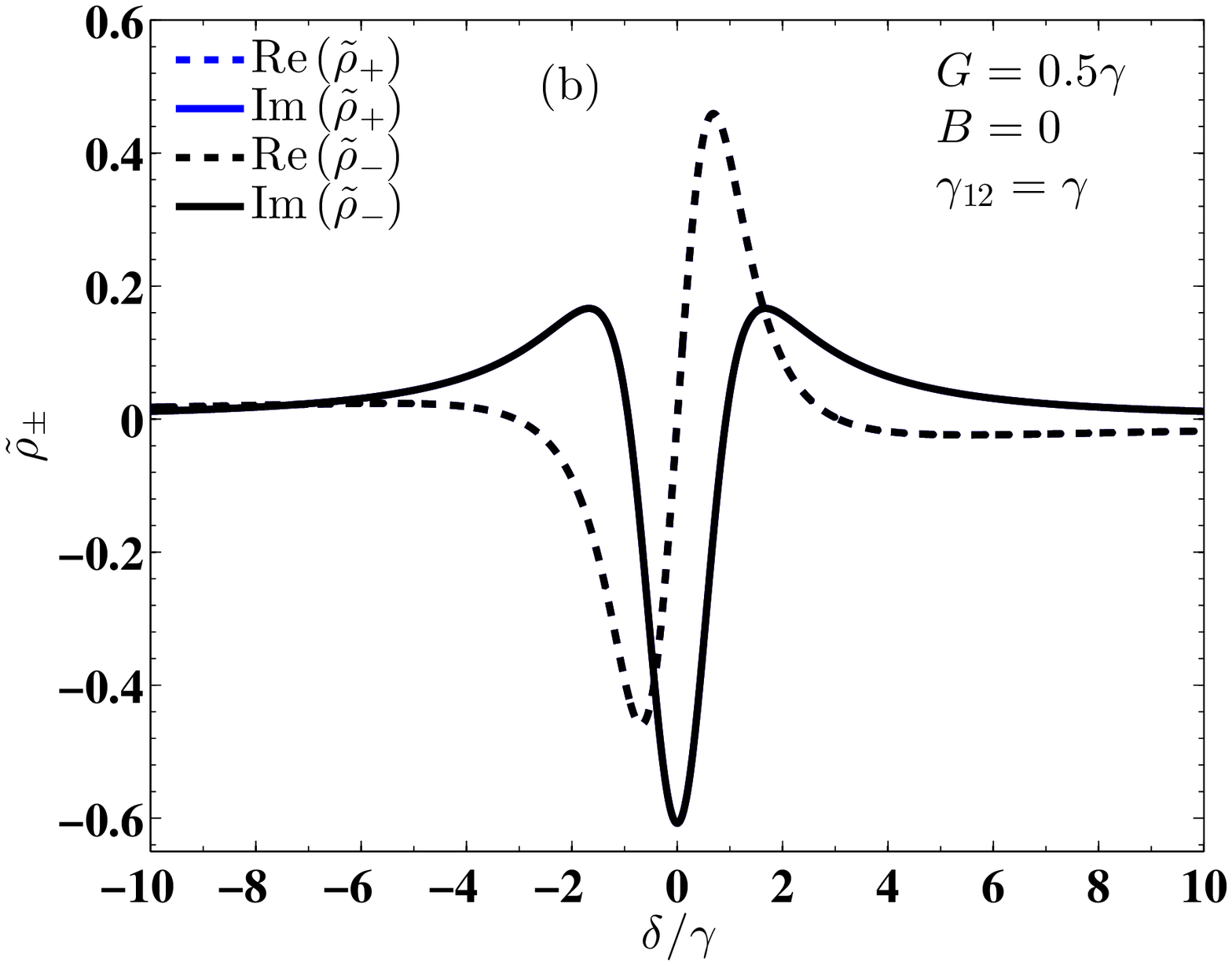}\\
\includegraphics[scale=0.4]{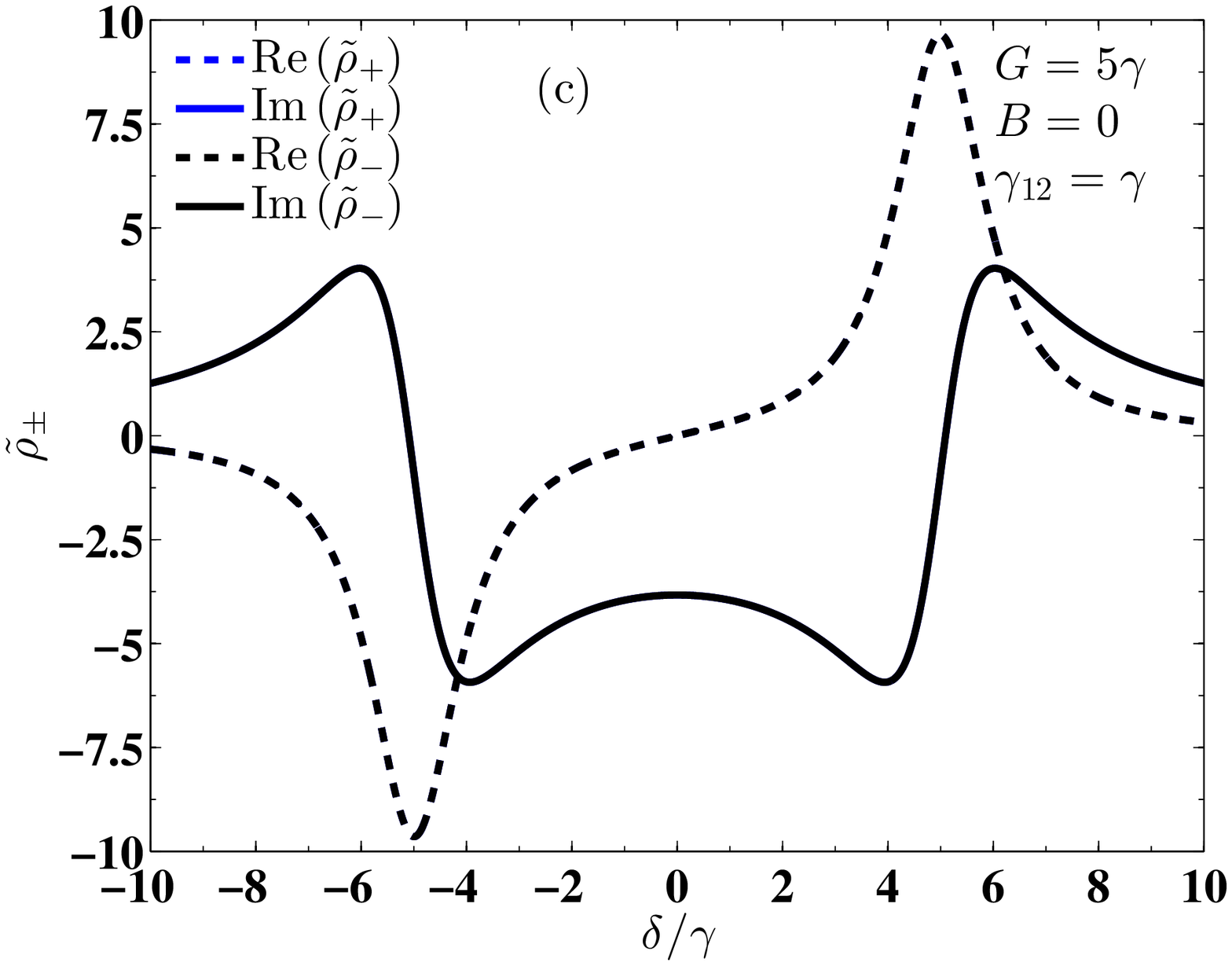}\\
\end{tabular}
\end{center}
\caption{(Color online) The variation of real (dotted line) and imaginary (solid line) parts of $\tilde{\rho}_{\pm}$ with the probe field detuning $\delta/\gamma$. The parameters used are (a) $G=0$, (b) $G=0.5\gamma$,
(c) $G=5\gamma$. Other common parameters are $B=0$, $\theta=0$, $\Delta=0$, $\tilde{\rho}_{00}^{(0)}=1$, $\tilde{\rho}_{11}^{(0)}=0$, $\tilde{\rho}_{22}^{(0)}=0$, $\tilde{\rho}_{33}^{(0)}=0$, $\gamma_{10} = \gamma_{20} =\gamma$, $\gamma_{30} = 0.1\gamma$, $\gamma_{13} = \gamma_{23} =0$,  $g_{1}=g_{2}=g_{1}^{\prime}=g_{2}^{\prime}=0.1\gamma$. In the absence of magnetic field, $\tilde{\rho}_{+}$
(blue line) overlaps over $\tilde{\rho}_{-}$ (black line). }
\label{fig3}
\end{figure}
We consider a situation where maximum VIC occurs
(i.e. $\theta=0$) in the system (for other cases see Appendix B). To obtain enhanced magneto-optical rotation, we would like to find a frequency domain, in which large circular birefringence occurs along with zero dichroism. Note that the MOR angle [Eq. (\ref{eq13})] is derived in the limit of no absorption of the probe field in the medium. This expression of MOR angle is however equally valid, if the absorption of the two components are equal and non-zero. In this case, the output probe field remains linearly polarized, though  with reduced intensity.

In Fig. \ref{fig3}, we present the dispersion and absorption profile of the polarization components. In absence of the control field [Fig. \ref{fig3}(a)], the system behaves as two two-level systems with degenerate excited states and exhibits absorption at resonance.  When the control field is switched on, large refractive index arises in the system at $\delta=\pm|G|$, along with zero absorption, whereas the absorption profile attains negative values at resonance \cite{scully1991,fleischhauer1992,rathe1993}, thanks to the coherence created by the control field between $|0\rangle$ and $|3\rangle$ [see Eq. (\ref{eq5})]. Fig. \ref{fig3}(b) elucidates the effect of control field for $G=0.5\gamma$. It can be inferred that refractive index attains high value at $\delta=\pm|G|$ where absorption is zero. When Rabi frequency of the control field is increased further to $G=5\gamma$, refractive index acquires much larger values at $\delta=\pm|G|$ as shown in Fig. \ref{fig3}(c). Thus, we can achieve larger values of refractive index without absorption by increasing Rabi frequency of the control field. It should be emphasized here that these profiles of $\sigma_\pm$ components are identical, as the magnetic field is not switched on yet. This clearly does not lead to any MOR, as the system is still isotropic with respect to the two polarization components.

\begin{figure}[!ht]
\begin{center}
\includegraphics[scale=0.4]{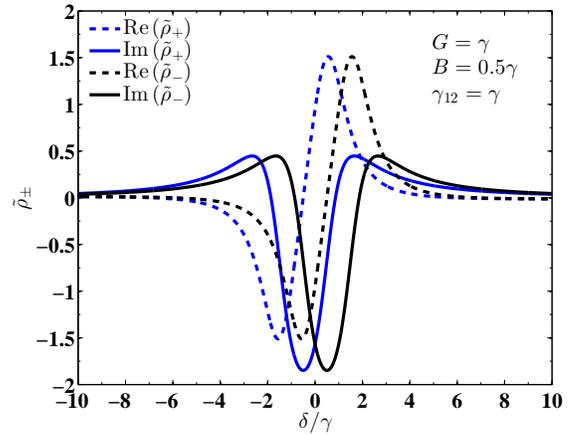}
\caption{(Color online) The evolution of real (dotted line) and imaginary (solid line) parts of $\tilde{\rho}_{\pm}$ with probe field detuning $\delta/\gamma$. The parameters used are $G=\gamma$, $B=0.5\gamma$ and the other parameters are the same as in Fig. \ref{fig3}.}
\label{fig4}
\end{center}
\end{figure}
\subsection{The effect of magnetic field}
Now let us elucidate the effect of the axial magnetic field. The applied magnetic field produces  Zeeman shift in each magnetic sublevel [see Fig. \ref{fig2}]. As a result, the resonance frequencies for the two circular components differ. Therefore, the right- and left-circularly polarized components exhibit different degrees of absorption and dispersion. We display in Fig. \ref{fig4} dispersion and absorption spectrum of the two polarization components in the presence of the magnetic field $\left(B=0.5\gamma\right)$. One can notice  from this figure that the peaks of absorption profiles occur at $\delta=\pm B$ whereas large refractive index with vanishing absorption can be obtained at $\delta=\pm\left(G\pm B\right)$. Moreover, the  magnetic field causes the absorption and dispersion profiles to be separated by an amount $2B$.

As discussed above, the magnetic field creates anisotropy in the medium while the control field provides large value of the refractive index with vanishing absorption. This anisotropy results in the difference in the refractive index (circular birefringence) for the circular components of the probe field which leads to  magneto-optical rotation  [Eq. (\ref{eq13})]. Besides this, magnetic field also induces difference in the absorption (circular dichroism) for right- and left-circularly polarized components. This difference in the imaginary parts of the susceptibilities causes ellipticity of the probe field. In this situation the complete description of the polarization state of the output probe field requires one to invoke Stoke's parameters \cite{born1999,dasgupta2003}.

\subsection{Magneto-optic rotation}
\begin{figure}[!ht]
\begin{center}
\begin{tabular}{c}
\includegraphics[scale=0.4]{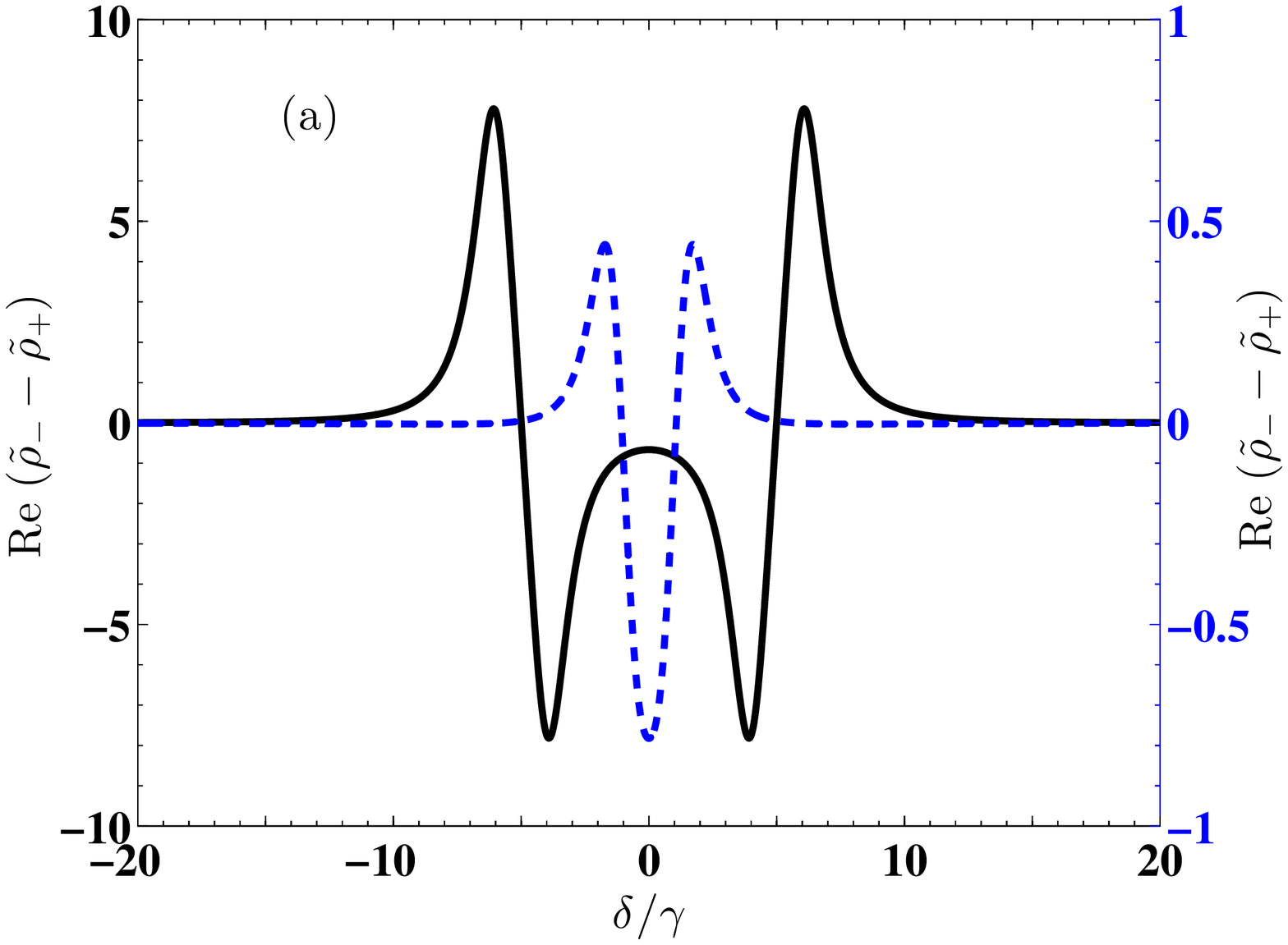}\\
\includegraphics[scale=0.4]{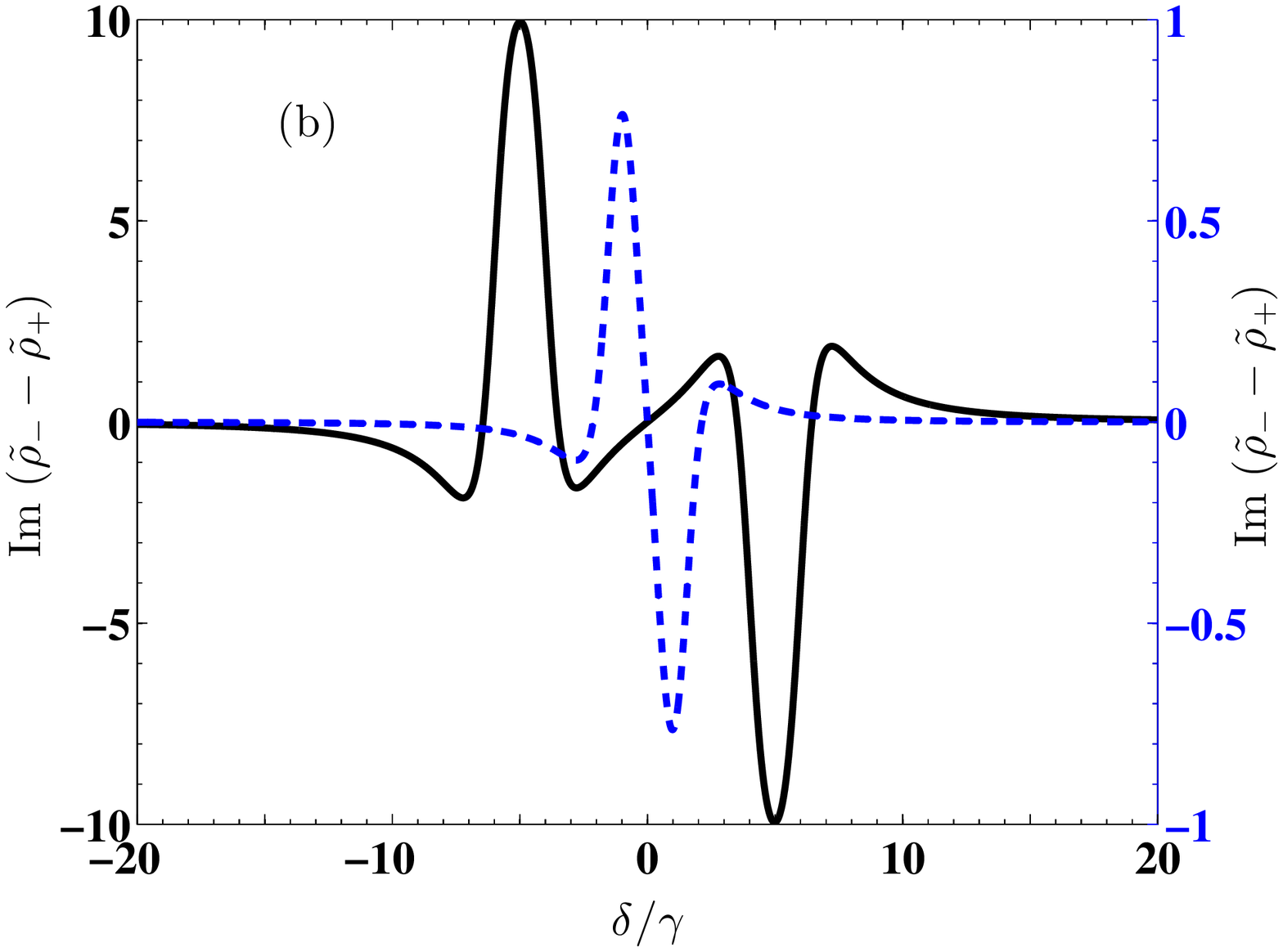}

\end{tabular}
\end{center}
\caption{(Color online) The variation of (a) the real parts and (b) the imaginary parts of the difference between $\tilde{\rho}_{-}$ and  $\tilde{\rho}_{+}$ with probe field detuning $\delta/\gamma$. The blue dotted line (blue ticks on the right on y-axis) is for $G=0.5\gamma$ whereas black solid line (black ticks on the left on y-axis) is for $G=5\gamma$.  We have chosen $B=\gamma$ and the other parameters are the same as used in Fig. \ref{fig3}.}
\label{fig5}
\end{figure}

We show in Fig. \ref{fig5}(a) that the  circular birefringence is enhanced at resonance for $G=0.5\gamma$ whereas for larger $G$ (e.g., for $G=5\gamma$), this occurs at $\delta=\pm\left(G\pm B\right)$. The large circular birefringence at resonance for $G=0.5\gamma$ can be attributed to the fact that the relevant dressed states are not well resolved for weaker field strength. On the other hand, by increasing the control field strength ($G=5\gamma$) the dressed states can be made split apart, so that the coherence at resonance becomes negligible. Large coherence and therefore large circular birefringence is achieved at  $\delta=\pm\left(G\pm B\right)$.  The enhancement in the birefringence is associated with vanishing circular dichroism  as depicted in Fig. \ref{fig5}(b).

\begin{figure}[!ht]
\begin{center}
\begin{tabular}{c}
\includegraphics[scale=0.4]{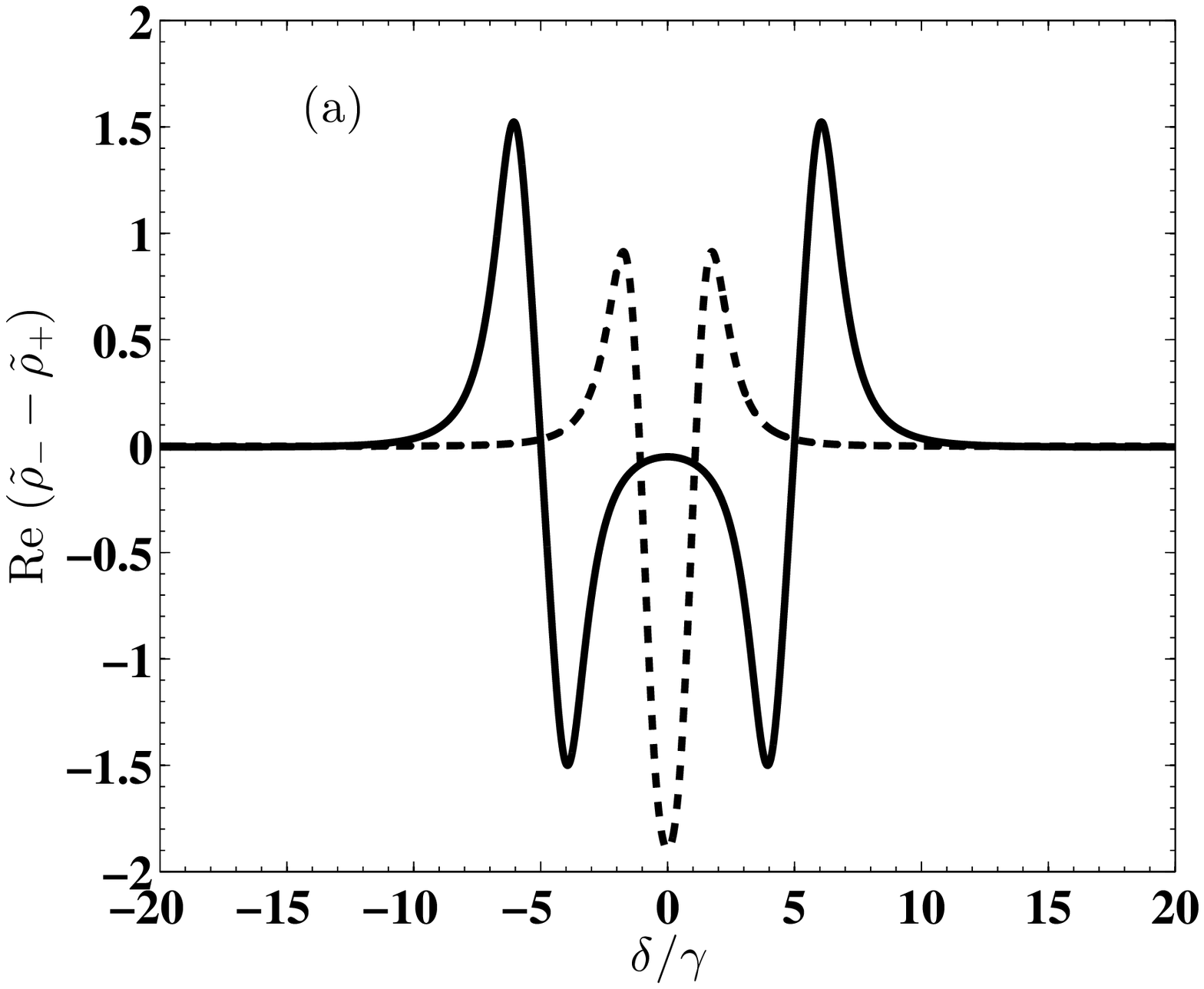} \\
\includegraphics[scale=0.4]{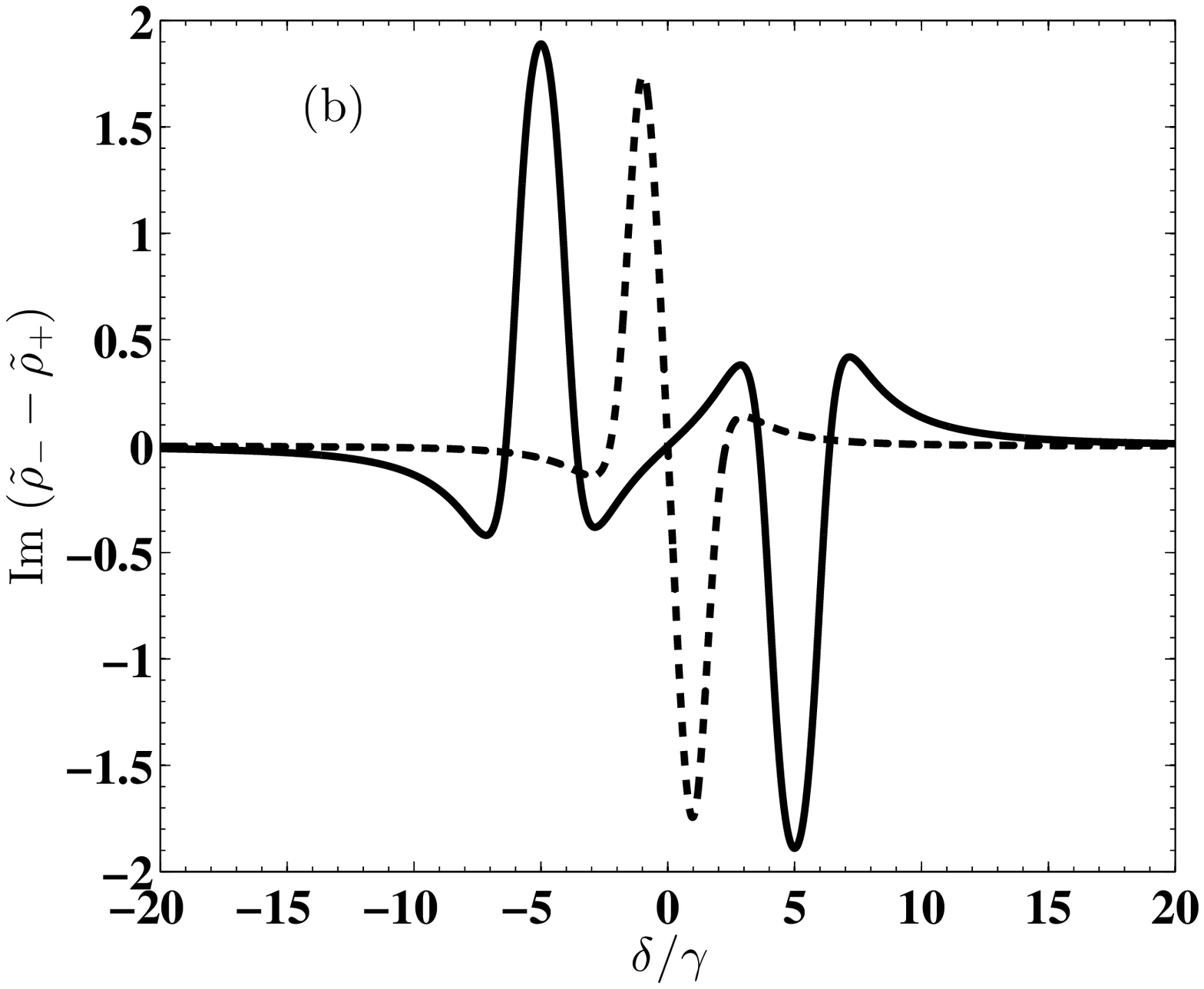}

\end{tabular}
\end{center}
\caption{ The variation of (a) Re$\left(\tilde{\rho}_{-}-\tilde{\rho}_{+}\right)$ and (b) Im$\left(\tilde{\rho}_{-}-\tilde{\rho}_{+}\right)$ with probe field detuning $\delta/\gamma$ for $G=\gamma$ and different magnetic fields $B=0.5\gamma$ (dotted line) and  $B=5\gamma$ (solid line). The other parameters are the same as used in Fig. \ref{fig3}. }
\label{fig6}
\end{figure}

Further, the separation between absorption and dispersion profiles of $\sigma_{\pm}$ components of the probe field increases with increase in the magnetic field. We display in Fig. \ref{fig6}  the real and imaginary parts of the difference between the coherences for right- and left-polarized components as a function of  probe field detuning at different magnetic fields. Fig. \ref{fig6}(a) shows that, when magnetic field is small, say $ B=0.5\gamma$, there is a large difference in refractive index with zero circular dichroism [Fig. \ref{fig6}(b)] at $\delta=0$. This is the most sought after feature to obtain large MOR angle. Note that as the magnetic field is increased, the circular birefringence decreases while the circular dichroism remains zero at resonance. This happens because of larger Zeeman shifts of the two excited states with $m_J=\pm 1$ and therefore, larger shifts in the absorption and dispersion profiles.
\begin{figure}[!ht]
\begin{center}
\begin{tabular}{c}
\includegraphics[scale=0.4]{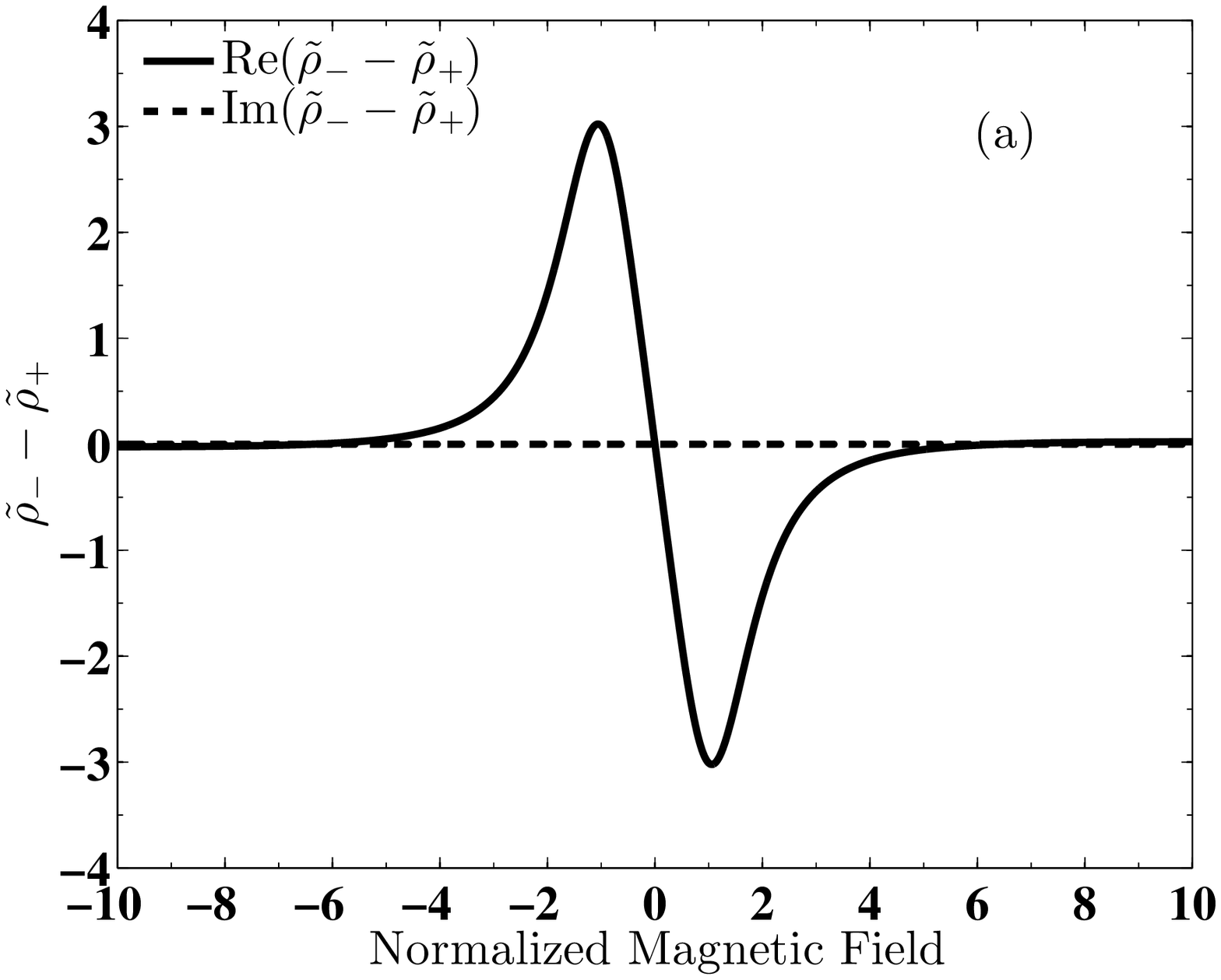}\\
\includegraphics[scale=0.4]{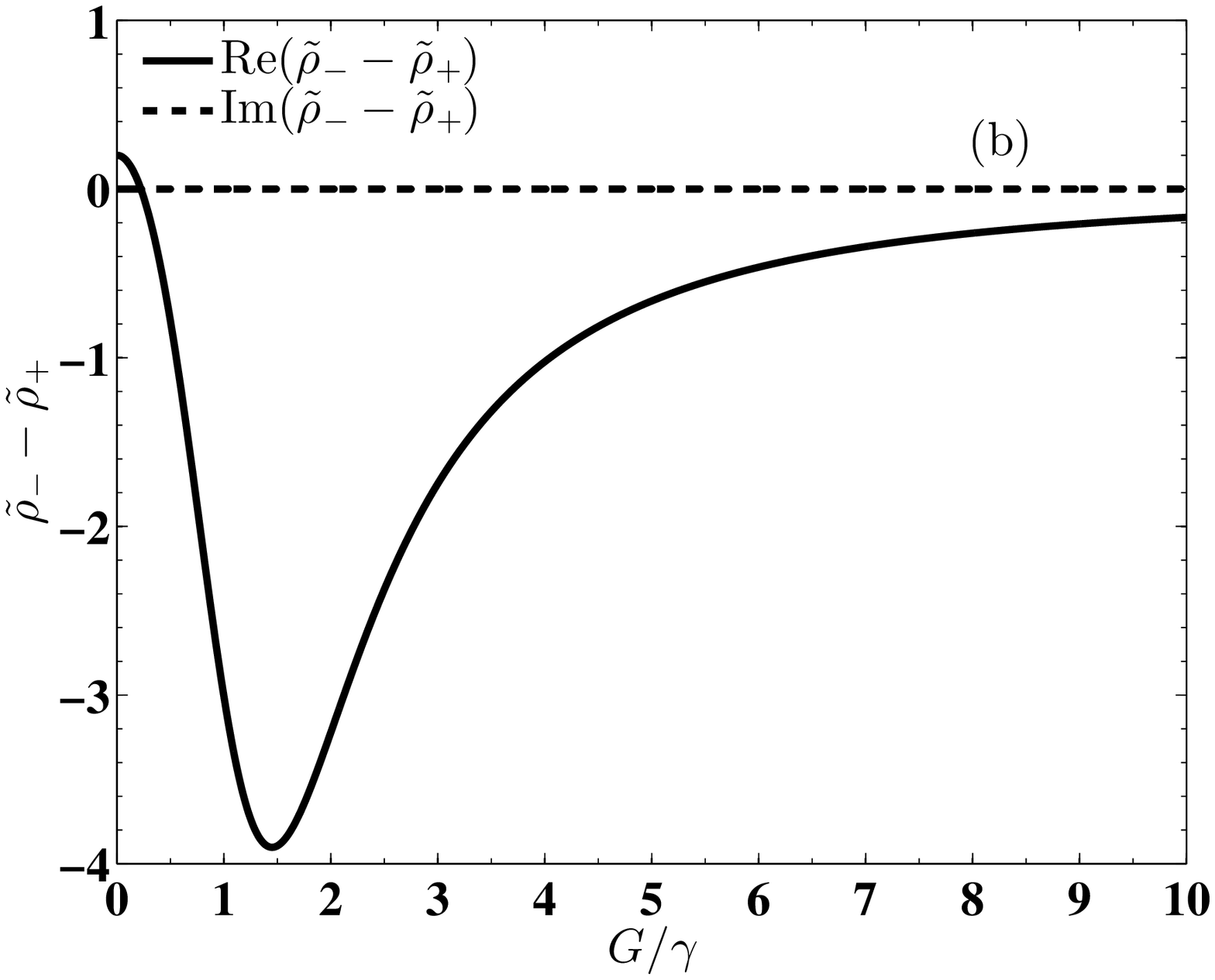}\\

\end{tabular}
\end{center}
\caption{ Variation of the real (solid line) and imaginary (dashed line) parts of the difference between $\tilde{\rho}_{-}$ and  $\tilde{\rho}_{+}$ at resonance ($\delta=0$) with (a) magnetic field $B$ for $G=\gamma$ and (b) Rabi frequency $G$ of the control field for $B=\gamma$. The other parameters are the same as used in Fig. \ref{fig3}.}
\label{fig7}
\end{figure}

We further demonstrate the effect of the magnetic and control field on circular birefringence and dichroism at probe resonance  in Fig. \ref{fig7}. The difference between the imaginary parts of $\tilde{\rho}_{\pm}$ at resonance remains zero as we increase the magnetic field. Thus, both $\sigma_{\pm}$ components of the probe field  equally couple to the two Zeeman-shifted atomic states throughout the medium  at resonance. On the other hand, the birefringence vanishes at resonance in the absence of magnetic field whereas it attains large values at $B=G=\gamma$ as illustrated in Fig. \ref{fig7}(a).  As shown in Fig. \ref{fig7}(b),   the absorptions for the two polarization components remain the same whereas birefringence achieves very high value for $G\approx B$ . It is to be noted that  birefringence is not zero at $G=0$ due to the presence of magnetic field.  Therefore, the condition $G=B$  at resonance is important for obtaining large MOR angle with zero circular dichroism.
\begin{figure}[!ht]
\begin{center}
\begin{tabular}{c}
\includegraphics[scale=0.4]{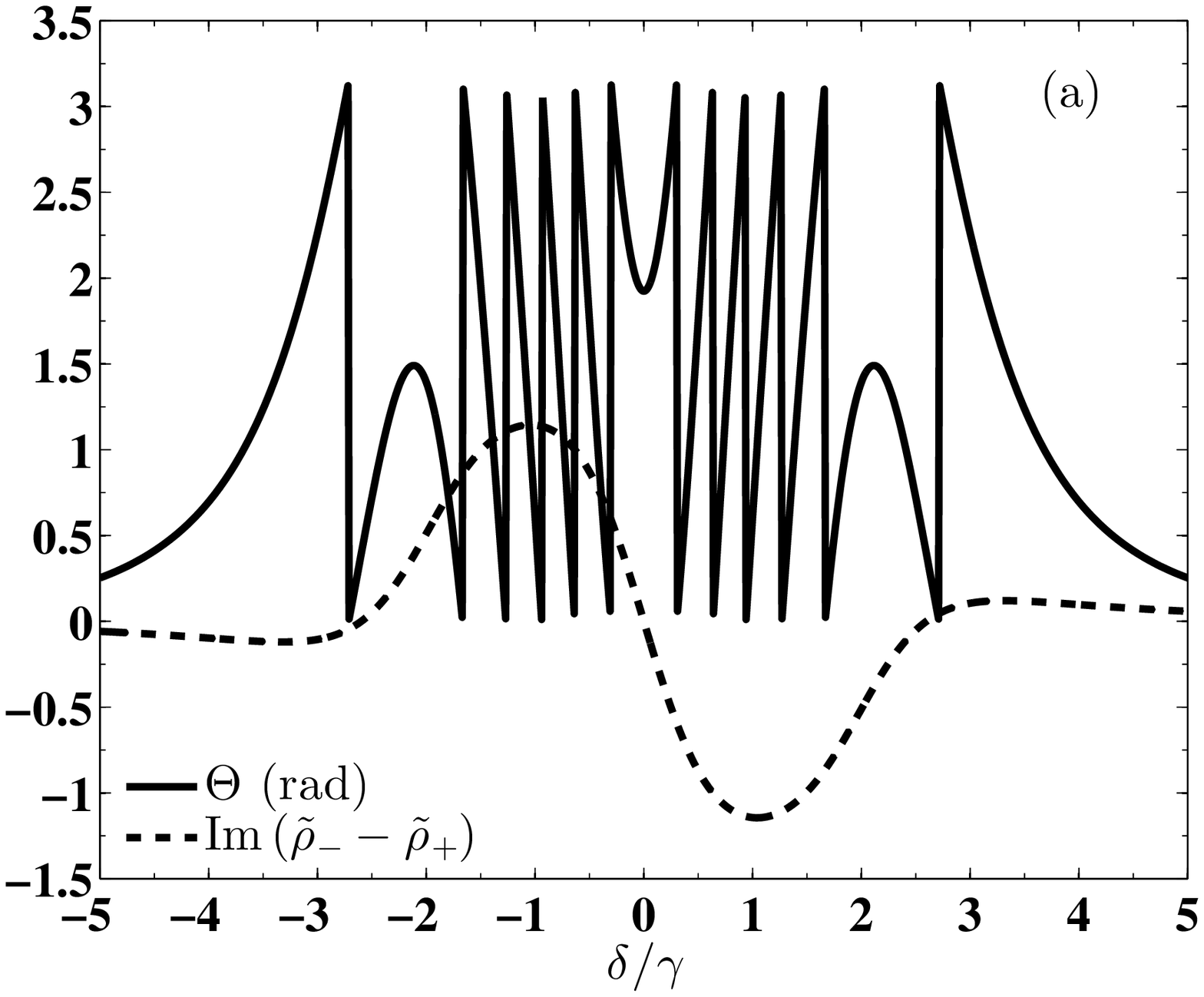}\\
\includegraphics[scale=0.4]{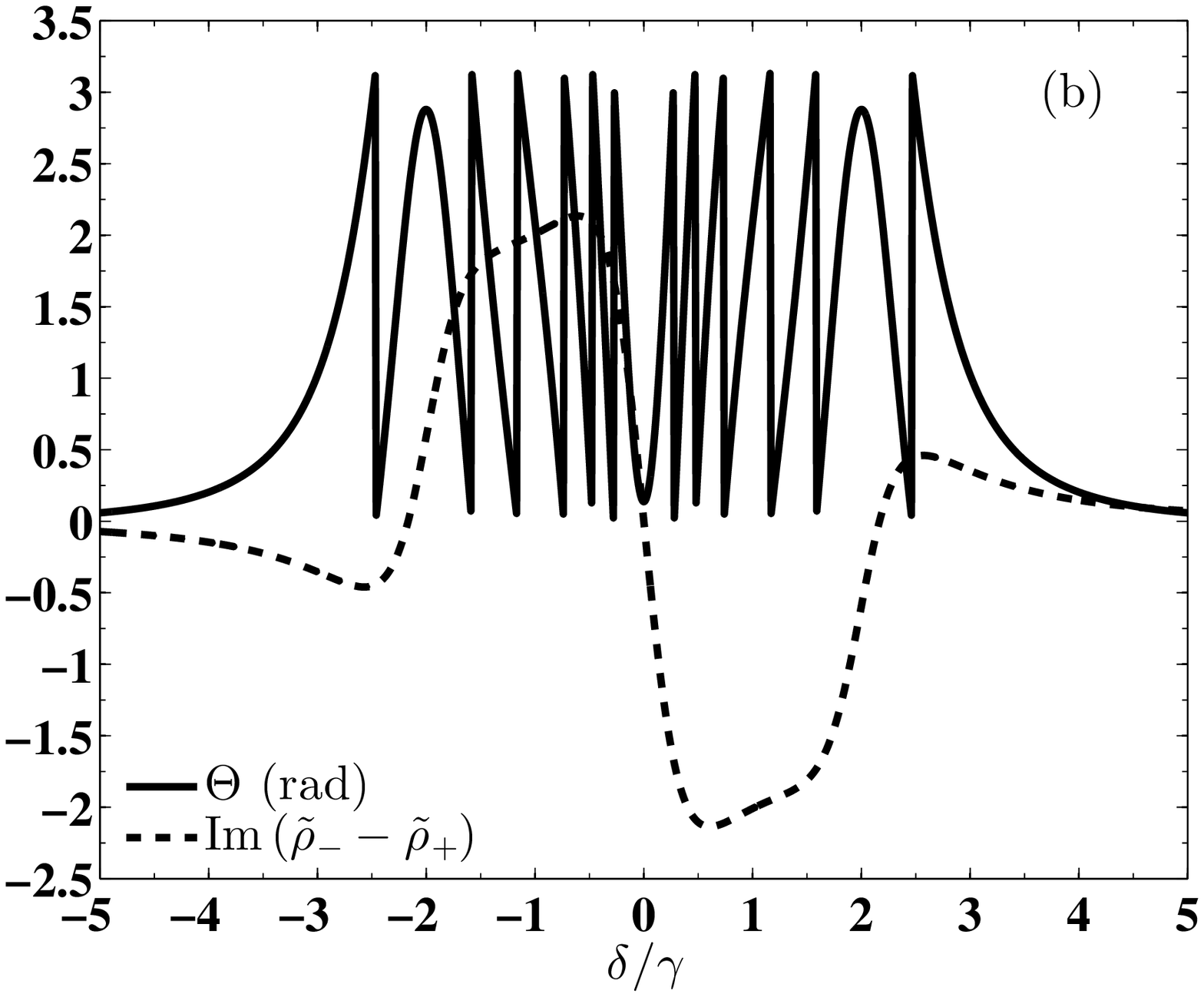}\\
\end{tabular}
\end{center}
\caption{ Variation of magneto-optical rotation angle $\Theta$ (in radians) (solid line) and circular dichroism (Im$\left(\tilde{\rho}_{-}-\tilde{\rho}_{+}\right)$) (dotted line) with detuning $\delta/\gamma$ of the probe field for (a) $\theta=0$,
and (b) $\theta=\pi/2$. We have chosen $G=B=\gamma$ and $\mathcal{C}= 3.5$ in Eq. (\ref{eq13}) and the other parameters are the same as  in Fig. \ref{fig3}. Here, MOR angle ($\Theta$) is obtained by applying modulo operator on $\mathcal{C}~\mbox{Re}\left(\tilde{\rho}_{-}-\tilde{\rho}_{+}\right)$ with $\pi$ \cite{rapidoscill}.}
\label{fig8}
\end{figure}

The above study suggests that the MOR angle can be enhanced at resonance without any ellipticity at the output. 
In Fig. \ref{fig8}, we show how the MOR angle and circular dichroism vary with probe field detuning $\left(\delta/\gamma\right)$, when the angle between the two dipole moments change between zero and $\pi/2$. It is evident in Fig. \ref{fig8}(a) that in presence of VIC ($\theta=0$), the MOR angle can be  large   at resonance, while in absence of VIC ($\theta=\pi/2$), this angle becomes much smaller at resonance [Fig. \ref{fig8}(b)]. It is to be noted that the  MOR angle becomes large also for $\theta=\pi/2$, but at detuned frequencies [Fig. \ref{fig8}(b)]. However, here, we focus on resonance, as the circular dichroism vanishes at resonance, thereby causing the polarization to remain linear at the output (Note that for large detuning the MOR angle is accompanied by large cicrcular dichroism, see Fig. \ref{fig8},  which results in the ellipticity of the output polarization). Therefore, it is worthwhile to explore the behavior of MOR at resonance, for which the parallel dipole moments give rise to large $\Theta$ in contrast to orthogonal case. We thus conclude that the enhancement in the angle of rotation of polarization at the output can be appraised as an effect of VIC and therefore can be chosen as a signature of VIC in the molecular systems.

\section{VIC based magnetometer}
 The preceding analysis delineates that the large MOR angle is a consequence of the cooperation of VIC and the control field. Such a large MOR angle due to the presence of VIC provides an extra knob to measure the weak magnetic field. Usually, a small magnetic field produces very small anisotropy in the system and hence a very small MOR angle which is difficult to detect. But, here we obtain  MOR angle as large as ($\sim 180^{\circ}$) even with small magnetic field. Thus, by measuring MOR angle we can estimate the feeble magnetic field. For example, at resonance for a magnetic field as low as $0.35\gamma$, a MOR angle as large as $111^{\circ}$ can be achieved for $\theta=0$ and $G=B$, whereas in the absence of VIC, this angle becomes much smaller $\sim 29^{\circ}$.  Further, the sensitivity of the detectable magnetic field is limited by particulars of experiment, for instance, by shot noise \cite{budker2000} and ac-Stark shifts \cite{fleischhauer2000}. However, the limitation due to ac-Stark shift can be compensated by adjusting the detuning between the hyperfine components of the upper level  in such a way that the light shifts cancel \cite{novikova2001}. Based on the shot noise limit, the sensitivity ($\delta B_{z}$) of the magnetic field along the direction of propagation is given by \cite{budker2000,hsu2008}
\begin{align}\label{eq14}
\delta B_{z}=\frac{1}{\sqrt{N_{ph}}}\left(\frac{\partial \Theta}{\partial B_{z}}\right)^{-1}
\end{align}
 where, $N_{ph}$ number photons detected and $\frac{\partial \Theta}{\partial B_{z}}$ is the slope of polarization rotation with the longitudinal magnetic field.
Here, the sensitivity of the proposed magnetometer using the parameters of \cite{grangier1987} turns out be of the order of 10$^{-14}\times\eta$ G/$\sqrt{\rm Hz}$ \cite{sensitivity}, where $\eta$ is the numeric value of $\gamma$. Thus, the magnetometer based on VIC can measure a very feeble magnetic field with very high sensitivity which is comparable to that reported in \cite{budker2000}. However, the magnetometers based on superconducting quantum interference devices (SQUID) and spin exchange relaxation-free (SERF) magnetometers have sensitivities of the order of 10$^{-15}$ T/$\sqrt{\rm Hz}$ and 10$^{-18}$ T/$\sqrt{\rm Hz}$ \cite{budker2007}. Thus, if the relaxation processes can be controlled in the proposed magnetometer, its sensitivity may be made comparable to these magnetometers. 

\section{Concluding Remarks}
In conclusions, we have shown how VIC can lead to the enhancement of MOR angle in a molecular system. We have
illustrated that the coupling between the degenerate excited states resulting from the interference between decay pathways can produce large value of MOR angle without ellipticity at the output. This provides a signature of VIC. We have analyzed how the combined effects of control and magnetic field in the presence of VIC can produce substantial MOR angle. We have identified a parameter regime of  these fields where resonant enhancement of birefringence and therefore of MOR angle is possible. Further, we have discussed how such a large MOR angle can act as the probe of weak magnetic field with large measurement sensitivity. MOR can happen without VIC, but VIC leads to the enhancement. So, experimentally, a proof of VIC can be established by showing this enhancement. To achieve this, one has to compare the results with and without VIC. To do an experiment without VIC is simple- just to choose two degenerate or nearly degenerate excited states with no restriction on the orthogonality of two transitions. To select two non-orthogonal transitions is the key to realize the VIC effects. Cold molecular systems will provide a useful platform in this regard. The current experimental efforts in producing cold molecules by photo \cite{jones2006}-or magneto-association \cite{kohler2006,chin2010} has mainly focused on the preparation of deeply bound molecules in absolute ro-vibrational singlet ground-state potential. Since such molecules are formed from ultracold atoms, they are by default translationally and rotationally cold. Therefore, such cold molecules offer experimental advantage in selecting the desired rotational and vibrational levels for realizing the discussed effects. Particularly, in the near-degenerate $\Lambda$-doubling case, deeply bound ground-state molecules in singlet state will be useful for inducing $^1\Sigma\leftrightarrow{}^1\Pi$ transitions. Moreover, in recent times, many groups have demonstrated direct laser cooling of molecules in singlet ground-state potential \cite{danzl2008,demille2008,carr2009}.

  In non-degenerate case, loosely bound ro-vibrational states in shallower ground-state triplet potential will be more useful, since one can then access the required upper levels from such ro-vibrational levels. Typically, the linewidth of molecular excited levels is of the order of 10 MHz. Therefore, one has to look for two excited vibrational levels having the same electronic angular momenta with vibrational spacing being less than or equal to the linewidth. With the rapid progress in the production of cold molecules in recent times \cite{krem2009,ni2008,danzl2010,carr2009}, the prospect for realization of VIC-assisted MOR in near future seems to be quite promising. An alternative approach to probe VIC using the MOR effect is to use two orthogonal photoassociative transitions from the collisional continuum of two ultracold atoms in the electronic ground state molecular potential.  

\begin{acknowledgments}
One of us (P.K.) would like to acknowledge the fruitful discussions with Mr. Subrata Saha during his visit to IACS.
\end{acknowledgments}

\appendix
\section{Density matrix equations and first order coherences}
The density matrix equations of motion in the rotating-wave approximation can be written as
\begin{widetext}
\vspace{-0.5cm}
\begin{align}\label{eq1a}
\dot{\tilde{\rho}}_{11} &= -\gamma_{10}\tilde{\rho}_{11}-\frac{\gamma_{12}}{2}\left(\tilde{\rho}_{12}+\tilde{\rho}_{21}\right)+i\left(g_{1}\tilde{\rho}_{01}-g_{1}^{\ast}\tilde{\rho}_{10}\right)\;,\nonumber\\
\dot{\tilde{\rho}}_{22} &= -\gamma_{20}\tilde{\rho}_{22}-\frac{\gamma_{12}}{2}\left(\tilde{\rho}_{12}+\tilde{\rho}_{21}\right)+i\left(g_{2}\tilde{\rho}_{02}-g_{2}^{\ast}\tilde{\rho}_{20}\right)\;,\nonumber\\
\dot{\tilde{\rho}}_{33} &=-\gamma_{30}\tilde{\rho}_{33}+i\left(G\tilde{\rho}_{03}-G^{\ast}\tilde{\rho}_{30}\right)\;,\nonumber\\
\dot{\tilde{\rho}}_{21} &= \left(-2iB-\Gamma_{21}\right)\tilde{\rho}_{21}-\frac{\gamma_{12}}{2}\left(\tilde{\rho}_{11}+\tilde{\rho}_{22}\right)+i\left(g_{2}\tilde{\rho}_{01}-g_{1}^{\ast}\tilde{\rho}_{20}\right)\;,\nonumber\\
\dot{\tilde{\rho}}_{10} &= \left(i\delta_{10}-\Gamma_{10}\right)\tilde{\rho}_{10}-\frac{\gamma_{12}}{2}\tilde{\rho}_{20}+ig_{1}\left(1-\tilde{\rho}_{22}-2\tilde{\rho}_{11}-\tilde{\rho}_{33}\right)-ig_{2}\tilde{\rho}_{12}-iG\tilde{\rho}_{13}\;,\nonumber\\
\dot{\tilde{\rho}}_{20} &= \left(i\delta_{20}-\Gamma_{20}\right)\tilde{\rho}_{20}-\frac{\gamma_{12}}{2}\tilde{\rho}_{10}+ig_{2}\left(1-\tilde{\rho}_{11}-2\tilde{\rho}_{22}-\tilde{\rho}_{33}\right)-ig_{1}\tilde{\rho}_{21}-iG\tilde{\rho}_{23}\;,\nonumber\\
\dot{\tilde{\rho}}_{30} &= \left(i\delta_{30}-\Gamma_{30}\right)\tilde{\rho}_{30}+iG\left(1-\tilde{\rho}_{11}-2\tilde{\rho}_{33}-\tilde{\rho}_{22}\right)-ig_{1}\tilde{\rho}_{31}-ig_{2}\tilde{\rho}_{32}\;,\nonumber\\
\dot{\tilde{\rho}}_{31} &= \left[i\left(\delta_{30}-\delta_{10}\right)-\Gamma_{31}\right]\tilde{\rho}_{31}-\frac{\gamma_{12}}{2}\tilde{\rho}_{32}+iG\tilde{\rho}_{01}-ig_{1}^{\ast}\tilde{\rho}_{30}\;,\nonumber\\
\dot{\tilde{\rho}}_{32} &= \left[i\left(\delta_{30}-\delta_{20}\right)-\Gamma_{32}\right]\tilde{\rho}_{32}-\frac{\gamma_{12}}{2}\tilde{\rho}_{31}+iG\tilde{\rho}_{02}-ig_{2}^{\ast}\tilde{\rho}_{30}\;.
\end{align}
\end{widetext}
The above density matrix elements obey the condition $\tilde{\rho}_{00}+\tilde{\rho}_{11}+\tilde{\rho}_{22}+\tilde{\rho}_{33}=1$ and $\tilde{\rho}_{ij}=\tilde{\rho}_{ji}^{\ast}$. Here, $\delta_{i0}=\delta+B=\omega_{p}-\omega_{i0}+B~$ is the detuning of the probe field from $|i\rangle\leftrightarrow|0\rangle~(i=1,2)$ transition, $\delta_{30}=\Delta=\omega_{c}-\omega_{30}$ is the detuning of the control field from $|3\rangle\leftrightarrow|0\rangle$ transition, the dephasing rate of coherence between the levels $|j\rangle$ and $|i\rangle$ is $\Gamma_{ij}=\frac{1}{2}\sum\limits_{k}\left(\gamma_{ki}+\gamma_{kj}\right)+\gamma_{coll}$, where $\gamma_{coll}$ is the collisional decay rate.  The highly oscillating terms in Eq. (\ref{eq1a}) has been neglected by using the transformations: $\rho_{10}=\tilde{\rho}_{10}e^{-i\omega_{p}t}$, $\rho_{20}=\tilde{\rho}_{20}e^{-i\omega_{p}t}$,   $\rho_{30}=\tilde{\rho}_{30}e^{-i\omega_{c}t}$, $\rho_{31}=\tilde{\rho}_{31}e^{-i\left(\omega_{c}-\omega_{p}\right)t}$, $\rho_{32}=\tilde{\rho}_{32}e^{-i\left(\omega_{c}-\omega_{p}\right)t}$ and $\rho_{ii}=\tilde{\rho}_{ii}$.

The steady state solutions of Eq. (\ref{eq1a}) can be found by using following perturbation expansion of the density matrix elements
\begin{equation}
\tilde{\rho}_{\alpha\beta}=\tilde{\rho}_{\alpha\beta}^{(0)}+g_{1}\tilde{\rho}_{\alpha\beta}^{\prime(+1)}+
+g_{2}\tilde{\rho}_{\alpha\beta}^{\prime(-1)}+c.c.
\label{eq2a}
\end{equation}
Thus, we obtain a set of algebraic equations of $\tilde{\rho}_{\alpha\beta}^{(n)}$. These equations can be solved for different values of $n$ to obtain following first coherence for $\sigma_{+}$ component of the probe field
\begin{widetext}
\begin{eqnarray}
\tilde{\rho}_{i0}^{\prime(+1)} &=& \frac{\left[\begin{array}{l}
-\left\{\Delta_{j0}\left(\Delta_{30}^{\ast}-\Delta_{i0}\right)\left(\Delta_{30}^{\ast}-\Delta_{j0}\right)+|G|^{2}\left(\Delta_{30}^{\ast}-\Delta_{i0}\right)+\Delta_{j0}\frac{\gamma_{ij}^{2}}{4}\right\}\left(\tilde{\rho}_{00}^{(0)}-\tilde{\rho}_{ii}^{(0)}\right)\\
-i\frac{\gamma_{ij}}{2}\left\{\left(\Delta_{30}^{\ast}-\Delta_{i0}\right)\left(\Delta_{30}^{\ast}-\Delta_{j0}\right)+|G|^{2}+\frac{\gamma_{ij}^{2}}{4}\right\}\tilde{\rho}_{ji}^{(0)}\\
+G\left\{\Delta_{j0}\left(\Delta_{30}^{\ast}-\Delta_{j0}\right)+|G|^{2}+\frac{\gamma_{ij}^{2}}{4}\right\}\tilde{\rho}_{03}^{(0)}\end{array}\right]}{\left[ \begin{array}{l}
\left(\Delta_{i0}\Delta_{j0}+\frac{\gamma_{ij}^{2}}{4}\right)\left\{\left(\Delta_{30}^{\ast}-\Delta_{i0}\right)\left(\Delta_{30}^{\ast}-\Delta_{j0}\right)\right\}\\
+|G|^{2}\left\{\Delta_{i0}\left(\Delta_{30}^{\ast}-\Delta_{i0}+\frac{\gamma_{ij}^{2}}{4}\right)+\Delta_{j0}\left(\Delta_{30}^{\ast}-\Delta_{j0}\right)+|G|^{2}+\frac{\gamma_{ij}^{2}}{4}\right\}\end{array}\right]}\label{eq3a}\\
\tilde{\rho}_{i0}^{\prime(-1)} &=& \frac{\left[\begin{array}{l}
\left\{\Delta_{j0}\left(\Delta_{30}^{\ast}-\Delta_{i0}\right)\left(\Delta_{30}^{\ast}-\Delta_{j0}\right)+|G|^{2}\left(\Delta_{30}^{\ast}-\Delta_{i0}\right)+\Delta_{j0}\frac{\gamma_{ij}^{2}}{4}\right\}\tilde{\rho}_{ij}^{(0)}\\
+\frac{i\gamma_{ij}}{2}\left\{\left(\Delta_{30}^{\ast}-\Delta_{i0}\right)\left(\Delta_{30}^{\ast}-\Delta_{j0}\right)+|G|^{2}+\frac{\gamma_{ij}^{2}}{4}\right\}\left(\tilde{\rho}_{00}^{(0)}-\tilde{\rho}_{jj}^{(0)}\right)\\
+\frac{iG\gamma_{ij}}{2}\left(\Delta_{i0}+\Delta_{j0}-\Delta_{30}^{\ast}\right)\tilde{\rho}_{03}^{(0)}\end{array}\right]}{\left[ \begin{array}{l}
\left(\Delta_{i0}\Delta_{j0}+\frac{\gamma_{ij}^{2}}{4}\right)\left\{\left(\Delta_{30}^{\ast}-\Delta_{i0}\right)\left(\Delta_{30}^{\ast}-\Delta_{j0}\right)\right\}\\
+|G|^{2}\left\{\Delta_{i0}\left(\Delta_{30}^{\ast}-\Delta_{i0}+\frac{\gamma_{ij}^{2}}{4}\right)+\Delta_{j0}\left(\Delta_{30}^{\ast}-\Delta_{j0}\right)+|G|^{2}+\frac{\gamma_{ij}^{2}}{4}\right\}\end{array}\right]}
\label{eq4a}
\end{eqnarray}
\end{widetext}
Here, $\Delta_{j0}=\delta_{j0}+i\Gamma_{j0}$ ($i,j=1,2$ and $i\neq j$), $\Delta_{30}=\delta_{30}+i\Gamma_{30}$.
\section{Nonparallel dipole arrangement}
\begin{figure}[h!]
\begin{center}
\includegraphics[scale=0.23]{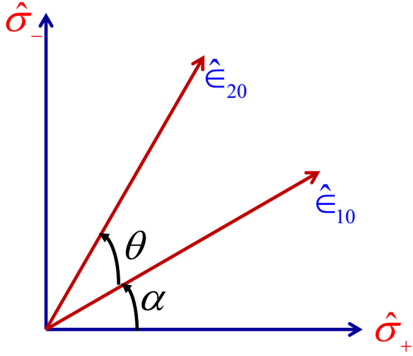} 
\caption{(Color online) Arrangement for non-parallel dipole moments. $\hat{\epsilon}_{10}$ and $\hat{\epsilon}_{20}$ are unit vectors along the direction of dipole moments, $\theta$ is the angle between dipole moments and $\alpha$ is the angle between $\hat{\epsilon}_{10}$ and $\hat{\sigma}_{+}$.  }
\label{fig9}
\end{center}
\end{figure}
The schematic of the nonparallel dipole arrangement is shown in Fig. \ref{fig9}. Here, $\hat{\epsilon}_{10}$ and $\hat{\epsilon}_{20}$ are unit vectors along the direction of dipole moments. Thus, we can write
\begin{align}\label{eqb1}
\vec{d}=d_{10}\hat{\epsilon}_{10}|1\rangle\langle 0|+d_{20}\hat{\epsilon}_{20}|2\rangle\langle 0|+h.c.\;,
\end{align}
 As $\theta$ is the angle between $\hat{\epsilon}_{10}$ ans $\hat{\epsilon}_{20}$ and $\alpha$  is the angle between $\hat{\epsilon}_{10}$ and $\hat{\sigma}_{+}$, so dipole moment unit vectors can be written as
\begin{align}
\hat{\epsilon}_{10} &= \cos(\alpha)\hat{\sigma}_{+} + \sin(\alpha)\hat{\sigma}_{-}\;,\\
\hat{\epsilon}_{20} &= \cos(\alpha+\theta)\hat{\sigma}_{+} + \sin(\alpha+\theta)\hat{\sigma}_{-}\;.
\end{align}
The interaction Hamiltonian can thus be given by
\begin{align}
H_{I}&=g_{+}^{nonparallel}|1\rangle\langle 0|+g_{-}^{nonparallel}|2\rangle\langle 0|+h.c.\;.
\end{align}
where the modified Rabi frequencies for the nonparallel arrangement takes the following form:-
\begin{align}
g_{+}^{nonparallel}&=\frac{1}{\sqrt{2}} \left(\sin(\alpha)+\sin(\alpha+\theta)\right)g_{+}\;,\\
g_{-}^{nonparallel}&=\frac{1}{\sqrt{2}}\left(\cos(\alpha)+\cos(\alpha+\theta)\right)g_{-}\;.
\end{align}
Thus, the nonparallel arrangement of the dipoles leads to the modification of the Rabi frequencies. The essence of VIC lies in the non-zero value of $\theta$. For a particular value of $\alpha$, the MOR angle can be calculated for non-zero $\theta $.

\bibliographystyle{elsarticl-num}

\begin{thebibliography}{60}
\bibitem{rapidoscill} The MOR angle ($\Theta$) in Fig. \ref{fig8} is obtained by applying the modulo operator on $\left[\mathcal{C}~\mbox{Re}\left(\tilde{\rho}_{-}-\tilde{\rho}_{+}\right)\right]$ with $\pi$. The rapid oscillations in $\Theta$ in  Fig. \ref{fig8} is a consequence of this modulo operation.
\end{thebibliography}

\end{document}